\pdfoutput=1
\documentclass[conference]{IEEEtran}
\usepackage{mathtools} 
\usepackage{amsthm, amssymb} 
\usepackage{bm} 
\usepackage{microtype} 
\usepackage[noadjust]{cite}
\usepackage[hidelinks, bookmarks = false]{hyperref} 
\usepackage[nameinlink]{cleveref} 
\usepackage{orcidlink} 

\usetikzlibrary{calc}

\hypersetup{hidelinks, pdfinfo = {Title = {On the Optimality of Coded Caching With Heterogeneous User Profiles}, Author = {Federico Brunero and Petros Elia}, Subject = {Information-Theoretic Caching}, Keywords = {Coded Caching, File Popularity, Index Coding, Information-Theoretic Converse, User Preferences}}}

\allowdisplaybreaks

\Crefformat{figure}{#2Fig.~#1#3}

\theoremstyle{plain}
\newtheorem{theorem}{Theorem}

\theoremstyle{definition}
\newtheorem{definition}{Definition}

\theoremstyle{remark}
\newtheorem{remark}{Remark}

\newcommand{\Nc}{N_{\textnormal{c}}}
\newcommand{\Nu}{N_{\textnormal{u}}}
\newcommand{\tc}{t_{\textnormal{c}}}
\newcommand{\tu}{t_{\textnormal{u}}}

\IEEEoverridecommandlockouts 

\title{On the Optimality of Coded Caching With Heterogeneous User Profiles\thanks{This work was supported by the European Research Council (ERC) through the EU Horizon 2020 Research and Innovation Program under Grant 725929 (Project DUALITY).}}

\author{\IEEEauthorblockN{Federico Brunero\textsuperscript{\orcidlink{0000-0002-6980-3827}} and Petros Elia\textsuperscript{\orcidlink{0000-0002-3531-120X}}}
\IEEEauthorblockA{Communication Systems Department, EURECOM, Sophia Antipolis, France\\
Email: \{brunero, elia\}@eurecom.fr}}

\begin{document}

\bstctlcite{IEEEexample:BSTcontrol}

\maketitle

\begin{abstract}
    In this paper, we consider a coded caching scenario where users have heterogeneous interests. Taking into consideration the system model originally proposed by Wang and Peleato, for which the end-receiving users are divided into groups according to their file preferences, we develop a novel information-theoretic converse on the optimal worst-case communication load under uncoded cache placement. Interestingly, the developed converse bound, jointly with one of the coded schemes proposed by Wang and Peleato, allows us to characterize the optimal worst-case communication load under uncoded prefetching within a constant multiplicative gap of $2$. Although we restrict the caching policy to be uncoded, our work improves the previously known order optimality results for the considered caching problem.
\end{abstract}


\section{Introduction}

With the introduction of video streaming platforms and cloud computing services, such as Netflix, Amazon Prime and AWS, we witnessed in the recent years a significant rise in the network traffic. On the one hand, the appearance of data-intensive network applications clearly entails more services at disposal of the users, with consequent benefits for the latter. On the other, communication networks are constantly put under pressure to deliver increasingly larger volumes of data in a timely and efficient manner. As a consequence, the explosion of network traffic has sparked much interest in developing new communication techniques and, to this end, much research focused on the pivotal role that caching will play in the future.

The idea of caching simply consists of exploiting low-cost memories at end-receiving users to diminish significantly the network load from a centralized server to its cache-aided receiving users. The real challenge is to intelligently design the \emph{placement phase} so as to minimize the volume of data that the server has to deliver during the \emph{delivery phase}. The seminal work in~\cite{MaddahAli2014FundamentalLimitsCaching} tackled this problem introducing the clever concept of coded caching, a major breakthrough which pushed even further the benefits of caching. More specifically, the Maddah-Ali and Niesen (MAN) coded scheme in~\cite{MaddahAli2014FundamentalLimitsCaching} shed light on the actual information-theoretic gains that caching can provide if the placement phase is carefully designed so that, during the delivery phase, coding techniques can be employed to align the interference patterns. Current research on coded caching spans several topics such as the interplay between multiple antennas and caching~\cite{Lampiris2018AddingTransmittersDramatically, Parrinello2020FundamentalLimitsCoded, Lampiris2022ResolvingFeedbackBottleneck, Toelli2020MultiAntennaInterference}, the construction of converse bounds~\cite{Wan2020IndexCodingApproach, Yu2018ExactRateMemory} and a variety of other scenarios~\cite{Hachem2017CodedCachingMulti, Serbetci2019MultiAccessCoded, Muralidhar2021MaddahAliNiesen, Brunero2021FundamentalLimitsCombinatorial}.

\subsection{Coded Caching With Hetereogeneous User Profiles}

Since the introduction of coded caching, many works studied several variations of the original information-theoretic system model. In particular, much research aimed at understanding how caching policies should reflect the possibility that, in real scenarios, contents may have different degrees of popularity and users may have diverging interests.

On the one hand, considering that traditional caching techniques heavily rely on the fact that some contents might be more popular than others, the works in~\cite{Niesen2017CodedCachingNonuniform, Zhang2018CodedCachingArbitrary} focused on the interplay between coded caching and file popularities.

On the other hand, some other works sought to explore the scenario where each user is not necessarily interested in the entire library of contents at the main server --- as instead was implicitly assumed in~\cite{MaddahAli2014FundamentalLimitsCaching}. For instance, the works in~\cite{Chang2019CodedCachingHeterogeneous, Chang2020CodedCachingTwo} explored, for the case of $K = 2$ users, the performance of selfish coded caching in the presence of heterogeneous user profiles or, equivalently, heterogeneous file demand sets (FDSs), proving that, for the instances proposed therein, unselfish caching policies can do better than selfish ones. Later, the work in~\cite{Brunero2021UnselfishCodedCaching} analyzed for a unified setting the interplay between selfish caching policies and coded caching, providing the meaningful conclusion that unselfish coded caching can be unboundedly better the selfish caching. Subsequently, the recent work in~\cite{Wan2022OptimalLoadMemory} characterized the optimal memory-load tradeoff for a scenario where users are interested in a limited set of contents which depends on the location of the users themselves. 

Recently, the authors in~\cite{Wang2019CodedCachingHeterogeneous, Zhang2020AverageRateCoded, Zhang2022CodedCachingHeterogeneous} considered coded caching with a very well-defined structure for the user profiles. Assuming that the files in the main library can be classified as either common files (files that can be requested by any user) or unique files (files that can be requested by groups of users only), the work in~\cite{Wang2019CodedCachingHeterogeneous} proposed three different coded schemes for such scenario, providing a related analysis for their peak load performance. Then, these three schemes were studied also in~\cite{Zhang2020AverageRateCoded} in terms of their average load performance, whereas the work in~\cite{Zhang2022CodedCachingHeterogeneous} provided, by means of a converse bound based on cut-set arguments, some order optimality results.

\subsection{Main Contribution}

Our work further explores the system model proposed in~\cite{Wang2019CodedCachingHeterogeneous}. In particular, taking advantage of the genie-aided converse bound idea from~\cite{Yu2018ExactRateMemory}, our main result is a lower bound on the optimal worst-case communication load under uncoded prefetching. Interestingly, the derived converse, together with an already existing achievable scheme from~\cite{Wang2019CodedCachingHeterogeneous}, allows us to characterize the memory-load tradeoff under uncoded placement within a constant multiplicative factor of $2$.

\subsection{Paper Outline}

The system model and related results are presented in \Cref{sec: System Model and Related Results}. \Cref{sec: Main Results} presents the information-theoretic converse and the order optimality result, whose proofs are provided in \Cref{sec: Converse Bound Proof} and \Cref{sec: Order Optimality Proof}, respectively. \Cref{sec: Conclusion} concludes the paper.

\subsection{Notation}

We denote by $\mathbb{Z}^{+}$ the set of positive integers. For $n \in \mathbb{Z}^{+}$, we define $[n] \coloneqq \{1, 2, \dots, n\}$. If $a, b \in \mathbb{Z}^{+}$ such that $a < b$, then $[a : b] \coloneqq \{a, a + 1, \dots, b - 1, b\}$. For sets we use calligraphic symbols, whereas for vectors we use bold symbols. Given a finite set $\mathcal{A}$, we denote by $|\mathcal{A}|$ its cardinality. We denote by $\binom{n}{k}$ the binomial coefficient and we let $\binom{n}{k} = 0$ whenever $n < 0$, $k < 0$ or $n < k$. For $n \in \mathbb{Z}^{+}$, we denote by $S_n$ the symmetric group defined over the set $[n]$.

\section{System Model and Related Results}\label{sec: System Model and Related Results}

We consider the coded caching setting where there is a single server connected to $K$ users through an error-free broadcast channel. The server has access to a central library that contains $N$ files of $B$ bits each. Each user in the system is equipped with a cache of size $MB$ bits (or, equivalently, $M$ files). According to the system model in~\cite{Wang2019CodedCachingHeterogeneous}, the $K$ users are split in $G$ groups, where each group consists of $K/G$ users sharing the same interests. Furthermore, the files in the library are divided in two categories, i.e., common files and unique files. There are $\Nc$ common files $\{W^{\text{c}}_{n} : n \in [\Nc]\}$, where each of them is of interest to every user in the system. Then, for each group $g \in [G]$, there are $\Nu$ files $\{W^{\text{u}, g}_{n} : n \in [\Nu]\}$, where each of them is of interest to the users belonging to the group $g \in [G]$ only. Assuming that $\{W^{\text{c}}_{n} : n \in [\Nc]\} \cap \{W^{\text{u}, g}_{n} : n \in [\Nu]\} = \emptyset$ for each $g \in [G]$, and that $\{W^{\text{u}, g_1}_{n} : n \in [\Nu]\} \cap \{W^{\text{u}, g_2}_{n} : n \in [\Nu]\} = \emptyset$ for each $g_1, g_2 \in [G]$ with $g_1 \neq g_2$, we have $N = \Nc + G\Nu$ files in total. Deviating from standard notation practices, we will use $W^{f_k, g(k)}_{d_k}$ to denote the file requested by user $k \in [K]$, where $f_k \in \{\text{c}, \text{u}\}$, $d_k \in [N_{f_k}]$ and $g(k)$ is an abuse of notation to denote the group which user $k$ belongs to, i.e., $g(k) \in [G]$ for each $k \in [K]$. We further assume that $W^{\text{c}, g(k)}_{d_k} = W^{\text{c}}_{d_k}$, since common files do not depend on the group $g \in [G]$. In addition, we let $\bm{d} = ((d_1, f_1), \dots, (d_K, f_K))$ be the demand vector and we denote by $\mathcal{D}$ the set of all possible demand vectors with distinct requested files, i.e., $W^{f_{k_1}, g(k_1)}_{d_{k_1}} \neq W^{f_{k_2}, g(k_2)}_{d_{k_2}}$ for each $k_1, k_2 \in [K]$ with $k_1 \neq k_2$. Finally, we assume $\Nc \geq K$ and $\Nu \geq K/G$.

The caching problem consists of two phases. During the placement phase, users have access to the main library, and so each user fills their cache using the library. Here, we focus on \emph{uncoded} caching policies according to the following definition.
\begin{definition}[Uncoded Prefetching]\label{def: Uncoded Placement Definition}
    A cache placement is \emph{uncoded} if each user $k \in [K]$ simply copies in their cache a total of (at most) $MB$ bits from the library. Consequently, the files are partitioned as
    \begin{align}
        W^{\text{c}}_n & = \{W^{\text{c}}_{n, \mathcal{T}} : \mathcal{T} \subseteq [K]\}, \quad \forall n \in [\Nc] \\
        W^{\text{u}, g}_n & = \{W^{\text{u}, g}_{n, \mathcal{T}} : \mathcal{T} \subseteq [K]\}, \quad \forall n \in [\Nu], \quad \forall g \in [G]
    \end{align}
    where $W^{\text{c}}_{n, \mathcal{T}}$ and $W^{\text{u}, g}_{n, \mathcal{T}}$ represent the bits of $W^{\text{c}}_{n}$ and $W^{\text{u}, g}_{n}$, respectively, which are cached only by the users in $\mathcal{T}$.
\end{definition}

During the delivery phase, the demand vector $\bm{d} = ((d_1, f_1), \dots, (d_K, f_K))$ is revealed to the server. Denoting by $\mathcal{X}$ the set of caching schemes with uncoded placement, the server transmits a message $X$ of $R(\bm{d}, \chi, M)B$ bits for a given demand $\bm{d} \in \mathcal{D}$, a given uncoded cache placement $\chi \in \mathcal{X}$ and some given memory $M$. The quantity $R(\bm{d}, \chi, M)$ is called \emph{load} and our goal is to characterize the optimal worst-case communication load under uncoded cache placement, namely, we aim to characterize the quantity given by
\begin{equation}
    R^{\star}(M) = \min_{\chi \in \mathcal{X}} \max_{\bm{d} \in \mathcal{D}} R(\bm{d}, \chi, M).
\end{equation}
In the following, the dependency on $M$ will be implied for the sake of simplicity.

\subsection{An Existing Achievable Scheme}\label{sec: An Existing Achievable Scheme}

The authors in~\cite{Wang2019CodedCachingHeterogeneous} proposed for the aforementioned setting a coded scheme --- referred to as Scheme 2 in~\cite{Wang2019CodedCachingHeterogeneous} --- which treats separately the caching and the delivery of common and unique files.

\subsubsection{Placement Phase}
First, the cache of each user is split in two parts for some $0 \leq \beta \leq 1$, so that $\beta M$ is the part of cache that is devoted to store common files and $(1 - \beta)M$ is the part of cache that is devoted to store unique files. Then, common files $\{W^{\text{c}}_{n} : n \in [\Nc]\}$ are stored across the $K$ users using the MAN cache placement with memory $\beta M$. Similarly, unique files $\{W^{\text{u}, g}_{n} : n \in [\Nu]\}$ are stored across the $K/G$ users in group $g \in [G]$ using the MAN algorithm with memory $(1 - \beta)M$.

\subsubsection{Delivery Phase}
It was shown in~\cite{Wang2019CodedCachingHeterogeneous} that, when there are $\alpha$ users per group requesting unique files, the optimal worst-case load can be upper bounded as
\begin{equation}\label{eqn: Achievable Load}
    R^{\star} \leq \min_{\beta} \max_{\alpha} R(\beta, \alpha)
\end{equation}
where $R(\beta, \alpha)$ is defined as
\begin{equation}\label{eqn: Term to Approximate}
    R(\beta, \alpha) \coloneqq \frac{\binom{K}{\tc + 1} - \binom{G\alpha}{\tc + 1}}{\binom{K}{\tc}} + G\frac{\binom{K/G}{\tu + 1} - \binom{K/G - \alpha}{\tu + 1}}{\binom{K/G}{\tu}}
\end{equation}
with $\tc \coloneqq K\beta M/\Nc$ and $\tu \coloneqq K(1 - \beta)M/G\Nu$.

Since the works in~\cite{Wang2019CodedCachingHeterogeneous, Zhang2022CodedCachingHeterogeneous} treated the variables $K$, $G$, $\Nc$, $\Nu$ and $t \coloneqq KM/N$ as continuous\footnote{Indeed, if the quantities $K$, $G$, $\Nc$ and $\Nu$ are large enough, the rounding errors due to integer effects during calculations can be neglected.}, we do the same here for the sake of simplicity. Further, we extend the Scheme 2 in~\cite{Wang2019CodedCachingHeterogeneous} to the entire memory regime $0 \leq M \leq \Nc + \Nu$, using the Gamma function whenever the binomial coefficients in~\eqref{eqn: Term to Approximate} have non-integer arguments.

\subsection{A Genie-Aided Converse Bound}\label{sec: Genie-Aided Approach}

We will provide our converse bound on the optimal worst-case load under uncoded prefetching using the genie-aided approach in~\cite{Yu2018ExactRateMemory}. Consider a demand vector $\bm{d} \in \mathcal{D}$ and let $\bm{u} = (u_1, \dots, u_K) \in S_K$ be a permutation of the set $[K]$. Denoting by $Z_k$ the cache content of user $k \in [K]$, we can construct a genie-aided user with the following cache content
\begin{equation}
    Z' = \left(Z_{u_k} \setminus \left(\bigcup_{i \in [k - 1]} Z_{u_i} \cup W^{f_{u_i}, g(u_i)}_{d_{u_i}} \right) : k \in [K] \right)
\end{equation}
which is enough for such genie-aided user to inductively decode all the requested files from $(X, Z')$. Consequently, the following
\begin{align}
    R(\bm{d}, \chi)B & \geq H(X) \\
                        & \geq H(X \mid Z') \\
                        & \geq I\left( \left\{W^{f_{u_k}, g(u_k)}_{d_{u_k}}\right\}_{k \in [K]} ; X \mid Z' \right) \\
                        & = H\left(\left\{W^{f_{u_k}, g(u_k)}_{d_{u_k}}\right\}_{k \in [K]} \mid Z' \right) \\
                        & = \sum_{k \in [K]} \sum_{\mathcal{T} \in ([K] \setminus \{u_1, \dots, u_k\})} \left\lvert W^{f_{u_k}, g(u_k)}_{d_{u_k}, \mathcal{T}} \right\rvert
\end{align}
holds, which means that we have the following lower bound
\begin{equation}\label{eqn: Lower Bound Inequality}
    R(\bm{d}, \chi) \geq \sum_{k \in [K]} \sum_{\mathcal{T} \in ([K] \setminus \{u_1, \dots, u_k\})} \frac{\left\lvert W^{f_{u_k}, g(u_k)}_{d_{u_k}, \mathcal{T}} \right\rvert}{B}
\end{equation}
on the communication load for a given\footnote{We recall that the dependency on $M$ is implied for the sake of simplicity.} $\bm{d} \in \mathcal{D}$ and $\chi \in \mathcal{X}$. Since it will be of use later, we define the following
\begin{equation}
    R_{\text{LB}}(\bm{d}, \bm{u}, \chi) \coloneqq \sum_{k \in [K]} \sum_{\mathcal{T} \in ([K] \setminus \{u_1, \dots, u_k\})} \frac{\left\lvert W^{f_{u_k}, g(u_k)}_{d_{u_k}, \mathcal{T}} \right\rvert}{B}.
\end{equation}

\section{Main Results}\label{sec: Main Results}

The first result provides a converse bound on the optimal worst-case load under uncoded prefetching. The proof is presented in \Cref{sec: Converse Bound Proof}.

\begin{theorem}\label{thm: Converse Bound Result}
    For the coded caching problem with heterogeneous user profiles presented in \Cref{sec: System Model and Related Results}, the optimal worst-case load under uncoded cache placement is lower bounded as
    \begin{equation}
        R^{\star} \geq \min_{\beta} \frac{1}{2}\left( \frac{\binom{K}{\tc + 1}}{\binom{K}{\tc}} + G \frac{\binom{K/G}{\tu + 1}}{\binom{K/G}{\tu}}\right)
    \end{equation}
    where $\tc = K\beta M/\Nc$ and $\tu = K(1 - \beta)M/G\Nu$.
\end{theorem}

If we compare the achievable performance in~\eqref{eqn: Achievable Load} with the converse in \Cref{thm: Converse Bound Result}, we can provide the following optimality result, whose proof is described in \Cref{sec: Order Optimality Proof}.

\begin{theorem}\label{thm: Order Optimality Result}
    The achievable load in~\eqref{eqn: Achievable Load} is order optimal within a multiplicative factor of $2$. 
\end{theorem}

\begin{remark}
    The result in \Cref{thm: Order Optimality Result} improves the previously known order optimality results presented in~\cite{Zhang2022CodedCachingHeterogeneous}. Indeed, even though the work in~\cite{Zhang2022CodedCachingHeterogeneous} provided a converse bound without constraining the placement to be uncoded, the smallest gap to optimality therein was a constant factor $8$ for the limited memory regime $N/K \leq M \leq N/2G$. Moreover, the achievable performance in~\eqref{eqn: Achievable Load} was shown to be within a multiplicative factor of $8 + 8K/G$ from optimal for the memory regime $G(\Nc + \Nu)/K \leq M \leq N/2G$. Here, although our converse holds under the assumption of uncoded placement, we provide a gap to optimality which is a constant multiplicative factor of $2$ for the entire\footnote{The bound in \Cref{thm: Converse Bound Result} becomes $0$ only when it holds $\tc = K$ and $\tu = K/G$ simultaneously. This happens when $\beta = \Nc/M$ and $(1 - \beta) = \Nu/M$, which implies $0 \leq M \leq \Nc + \Nu$. In addition, we recall that the Scheme 2 in~\cite{Wang2019CodedCachingHeterogeneous} is extended to the entire memory regime $0 \leq M \leq \Nc + \Nu$.} memory regime $0 \leq M \leq \Nc + \Nu$.
\end{remark}

\section{Converse Bound Proof}\label{sec: Converse Bound Proof}

We recall that our goal is to lower bound the quantity
\begin{equation}
    R^\star = \min_{\chi \in \mathcal{X}} \max_{\bm{d} \in \mathcal{D}} R(\bm{d}, \chi).
\end{equation}
where again the dependency on $M$ is implied to simplify the notation. Denote by $\mathcal{D}_{\text{c}}$ the subset of $\mathcal{D}$ that contains all demands for which users make requests only from common files, which implies $\bm{d} = ((d_1, \text{c}), \dots, (d_K, \text{c}))$ for each $\bm{d} \in \mathcal{D}_{\text{c}}$. Similarly, denote by $\mathcal{D}_{\text{u}}$ the subset of $\mathcal{D}$ for which users make requests only from unique files, which implies $\bm{d} = ((d_1, \text{u}), \dots, (d_K, \text{u}))$ for each $\bm{d} \in \mathcal{D}_{\text{u}}$. One can see that $|\mathcal{D}_{\text{c}}| = \binom{\Nc}{K}K!$ and $|\mathcal{D}_{\text{u}}| = \left(\binom{\Nu}{K/G}(K/G)!\right)^G$. Then, we proceed to lower bound the optimal worst-case load as follows
\begin{align}
    R^\star & = \min_{\chi \in \mathcal{X}} \max_{\bm{d} \in \mathcal{D}} R(\bm{d}, \chi) \\
               & \geq \min_{\chi \in \mathcal{X}} \max\left( \max_{\bm{d} \in \mathcal{D}_{\text{c}}} R(\bm{d}, \chi), \max_{\bm{d} \in \mathcal{D}_{\text{u}}} R(\bm{d}, \chi) \right) \label{eqn: Demand Constraint} \\
               & \geq \min_{\chi \in \mathcal{X}} \frac{1}{2}\left( \max_{\bm{d} \in \mathcal{D}_{\text{c}}} R(\bm{d}, \chi) + \max_{\bm{d} \in \mathcal{D}_{\text{u}}} R(\bm{d}, \chi) \right) \label{eqn: Average Lower Bound 1} \\
               & \geq \min_{\chi \in \mathcal{X}} \frac{1}{2}\left( \frac{1}{|\mathcal{D}_{\text{c}}|} \sum_{\bm{d} \in \mathcal{D}_{\text{c}}} R(\bm{d}, \chi) + \frac{1}{|\mathcal{D}_{\text{u}}|} \sum_{\bm{d} \in \mathcal{D}_{\text{u}}} R(\bm{d}, \chi) \right) \label{eqn: Average Lower Bound 2} \\
               & = \min_{\chi \in \mathcal{X}} \frac{1}{2}\left( R_{\text{c}}(\chi) + R_{\text{u}}(\chi) \right)
\end{align}
where $R_{\text{c}}(\chi)$ and $R_{\text{u}}(\chi)$ are defined as
\begin{align}
    R_{\text{c}}(\chi) & \coloneqq \frac{1}{|\mathcal{D}_{\text{c}}|} \sum_{\bm{d} \in \mathcal{D}_{\text{c}}} R(\bm{d}, \chi) \\
    R_{\text{u}}(\chi) & \coloneqq \frac{1}{|\mathcal{D}_{\text{u}}|} \sum_{\bm{d} \in \mathcal{D}_{\text{u}}} R(\bm{d}, \chi).
\end{align}
Notice that~\eqref{eqn: Demand Constraint} holds because $(\mathcal{D}_{\text{c}} \cup \mathcal{D}_{\text{u}}) \subset \mathcal{D}$, whereas both~\eqref{eqn: Average Lower Bound 1} and~\eqref{eqn: Average Lower Bound 2} follow from the fact that the maximum can be lower bounded by the average.

We proceed now to lower bound separately $R_{\text{c}}(\chi)$ and $R_{\text{u}}(\chi)$ by means of the genie-aided approach in \Cref{sec: Genie-Aided Approach}.

\subsection{Lower Bounding \texorpdfstring{$R_{\textnormal{c}}(\chi)$}{Rc(chi)}}
As we observed in \Cref{sec: Genie-Aided Approach}, the communication load can be lower bounded, for a given demand $\bm{d}$ and a given caching scheme $\chi$, as in~\eqref{eqn: Lower Bound Inequality}. Hence, if we construct the inequality in~\eqref{eqn: Lower Bound Inequality} for each demand $\bm{d} \in \mathcal{D}_{\text{c}}$ and for each permutation of users $\bm{u} \in S_K$, and then we sum together all such inequalities, we obtain
\begin{equation}
    K! \sum_{\bm{d} \in \mathcal{D}_{\text{c}}} R(\bm{d}, \chi) \geq \sum_{(\bm{d}, \bm{u}) \in (\mathcal{D}_{\text{c}}, S_K)} R_{\text{LB}}(\bm{d}, \bm{u}, \chi)
\end{equation}
which can be further rewritten as
\begin{equation}\label{eqn: General Inequality 1}
    R_{\text{c}}(\chi) \geq \frac{1}{K! |\mathcal{D}_{\text{c}}|} \sum_{(\bm{d}, \bm{u}) \in (\mathcal{D}_{\text{c}}, S_K)} R_{\text{LB}}(\bm{d}, \bm{u}, \chi)
\end{equation}
recalling that $\bm{d} = ((d_1, \text{c}), \dots, (d_K, \text{c}))$ for each $\bm{d} \in \mathcal{D}_{\text{c}}$ and that $W^{\text{c}, g}_{n} = W^{\text{c}}_{n}$ for each $n \in [\Nc]$ and for each $g \in [G]$. Now, towards simplifying the expression in~\eqref{eqn: General Inequality 1}, we proceed by counting how many times each subfile $W^{\text{c}}_{n, \mathcal{T}}$ --- for any given $n \in [\Nc]$ and $\mathcal{T} \subseteq [K]$ --- appears in the RHS of~\eqref{eqn: General Inequality 1}.

First, we focus on the subfile $W^{\text{c}}_{n, \mathcal{T}}$ for some $n \in [\Nc]$ and $\mathcal{T} \subseteq [K]$ such that $|\mathcal{T}| = t'$ with $t' \in [0 : K]$. Next, we denote by $\mathcal{D}_{\text{c}, n, k}$ the subset of demands in $\mathcal{D}_{\text{c}}$ for which the file $W^{\text{c}}_{n}$ is requested by some specific user $k \in ([K] \setminus \mathcal{T})$. We can see that $|\mathcal{D}_{\text{c}, n, k}| = \binom{\Nc}{K}K!/\Nc = |\mathcal{D}_{\text{c}}|/\Nc$. Then, we observe that, for each $\bm{d} \in \mathcal{D}_{\text{c}, n, k}$, all permutations of users $\bm{u} \in S_K$ are considered. Nevertheless, we can notice from the construction of~\eqref{eqn: Lower Bound Inequality} that, for each $\bm{d} \in \mathcal{D}_{\text{c}, n, k}$, the subfile $W^{\text{c}}_{n, \mathcal{T}}$ appears in the RHS of~\eqref{eqn: General Inequality 1} only for those permutations of users where $k$ appears before the elements from the set $\mathcal{T}$ in the permutation vector $\bm{u}$. Since there is a total of $(K - 1 - t')! t'! \binom{K}{t' + 1}$ such vectors, we can conclude that the subfile $W^{\text{c}}_{n, \mathcal{T}}$ appears in the RHS of~\eqref{eqn: General Inequality 1} a total of $|\mathcal{D}_{\text{c}}| (K - 1 - t')! t'! \binom{K}{t' + 1}/\Nc$ times when we consider the demands in $\mathcal{D}_{\text{c}, n, k}$ only. Then, since the reasoning above holds for each user $k \in ([K] \setminus \mathcal{T})$, we can conclude that the subfile $W^{\text{c}}_{n, \mathcal{T}}$ appears in the RHS of~\eqref{eqn: General Inequality 1} a total of $|\mathcal{D}_{\text{c}}| (K - t')! t'! \binom{K}{t' + 1}/\Nc$ times. Moreover, we considered a generic subfile $W^{\text{c}}_{n, \mathcal{T}}$, so the above holds for any $n \in [\Nc]$ and for any $\mathcal{T} \subseteq [K]$. Therefore, we can rewrite the RHS of~\eqref{eqn: General Inequality 1} as
\begin{equation}
    \frac{1}{K! |\mathcal{D}_{\text{c}}|} \sum_{t' \in [0 : K]} |\mathcal{D}_{\text{c}}| (K - t')! t'! \binom{K}{t' + 1} x^{\text{c}}_{t'}
\end{equation}
where $x^{\text{c}}_{t'}$ is defined as
\begin{equation}
    0 \leq x^{\text{c}}_{t'} \coloneqq \sum_{n \in [\Nc]} \sum_{\mathcal{T} \subseteq [K] : |\mathcal{T}| = t'} \frac{\left|W^{\text{c}}_{n, \mathcal{T}}\right|}{B\Nc}.
\end{equation}
After some algebraic manipulations, we can rewrite~\eqref{eqn: General Inequality 1} as
\begin{equation}
    R_{\text{c}}(\chi) \geq \sum_{t' \in [0 : K]} f_{\text{c}}(t') x^{\text{c}}_{t'}
\end{equation}
where $f_{\text{c}}(t')$ is defined as
\begin{equation}
    f_{\text{c}}(t') \coloneqq \frac{\binom{K}{t' + 1}}{\binom{K}{t'}}.
\end{equation}

\subsection{Lower Bounding \texorpdfstring{$R_{\textnormal{u}}(\chi)$}{Ru(chi)}}
Applying as before the genie-aided approach from \Cref{sec: Genie-Aided Approach}, we obtain the following inequality
\begin{equation}\label{eqn: General Inequality 2}
    R_{\text{u}}(\chi) \geq \frac{1}{K! |\mathcal{D}_{\text{u}}|} \sum_{(\bm{d}, \bm{u}) \in (\mathcal{D}_{\text{u}}, S_K)} R_{\text{LB}}(\bm{d}, \bm{u}, \chi)
\end{equation}
recalling that now $\bm{d} = ((d_1, \text{u}), \dots, (d_K, \text{u}))$ for each $\bm{d} \in \mathcal{D}_{\text{u}}$. Once again, towards simplifying the expression in~\eqref{eqn: General Inequality 2}, we proceed by counting how many times each subfile $W^{\text{u}, g}_{n, \mathcal{T}}$ --- for any given $n \in [\Nu]$, $g \in [G]$ and $\mathcal{T} \subseteq [K]$ --- appears in the RHS of~\eqref{eqn: General Inequality 2}.

First, we focus on the subfile $W^{\text{u}, g}_{n, \mathcal{T}_i}$ for a given $g \in [G]$, $n \in [\Nu]$ and $\mathcal{T}_i \subseteq [K]$ with $|\mathcal{T}_i| = t'$ for some $t' \in [0 : K]$, where $i$ denotes the number of users from group $g$ that appear in the set $\mathcal{T}_i$, namely, $i = |\{k \in \mathcal{T}_i : g(k) = g\}|$. Then, we let $k$ be one of the $(K/G - i)$ users from group $g$ that do not appear in $\mathcal{T}_i$ and we further assume that the file $W^{\text{u}, g}_{n}$ is requested by such user $k$. If we denote by $\mathcal{D}^{g}_{\text{u}, n, k}$ the subset of demands in $\mathcal{D}_{\text{u}}$ for which the file $W^{\text{u}, g}_{n}$ is requested by this user $k$, we can see that $|\mathcal{D}^{g}_{\text{u}, n, k}| = \left( \binom{\Nu}{K/G}(K/G)!\right)^G/\Nu = |\mathcal{D}_{\text{u}}|/\Nu$. In addition, for each $\bm{d} \in \mathcal{D}^{g}_{\text{u}, n, k}$, all permutations $\bm{u} \in S_K$ are considered. Nevertheless, as already observed, the subfile $W^{\text{u}, g}_{n, \mathcal{T}_i}$ appears in the RHS of~\eqref{eqn: General Inequality 2}, for each $\bm{d} \in \mathcal{D}^{g}_{\text{u}, n, k}$, only when $k$ is located before the elements from $\mathcal{T}_i$ in the permutation vector $\bm{u}$. Since there is a total of $(K - 1 - t')!t'!\binom{K}{t' + 1}$ such vectors, the subfile $W^{\text{u}, g}_{n, \mathcal{T}_i}$ appears $|\mathcal{D}_{\text{u}}|(K - 1 - t')!t'!\binom{K}{t' + 1}/\Nu$ times in the RHS of~\eqref{eqn: General Inequality 2} when only the demands $\mathcal{D}^{g}_{\text{u}, n, k}$ are considered. The same reasoning holds for any of the $(K/G - i)$ users from group $g$ not appearing in $\mathcal{T}_i$, so the subfile $W^{\text{u}, g}_{n, \mathcal{T}_i}$ appears $|\mathcal{D}_{\text{u}}| (K/G - i) (K - 1 - t')!t'!\binom{K}{t' + 1}/\Nu$ in the RHS of~\eqref{eqn: General Inequality 2}. In addition, since this reasoning holds for each $g \in [G]$, $n \in [\Nu]$ and $\mathcal{T}_i \subseteq [K]$ where $i \in [\max(0, |\mathcal{T}_i| - K + K/G), \min(|\mathcal{T}_i|, K/G)]$, after some algebraic manipulations we can rewrite the RHS of~\eqref{eqn: General Inequality 2} as
\begin{equation}
    \sum_{t' \in [0 : K]} \sum_{i = \max(0, t' - K + K/G)}^{\min(t', K/G)} \frac{(K/G - i)}{t' + 1} x^{\text{u}}_{t', i}
\end{equation}
where $x^{\text{u}}_{t', i}$ is defined as
\begin{equation}
    0 \leq x^{\text{u}}_{t', i} \coloneqq \sum_{g \in [G]} \sum_{n \in [\Nu]} \sum_{\substack{\mathcal{T}_i \subseteq [K] : |\mathcal{T}_i| = t',\\ |\{k \in \mathcal{T}_i: g(k) = g\}| = i}} \frac{\left|W^{\text{u}, g}_{n, \mathcal{T}_i}\right|}{B\Nu}.
\end{equation}
We can further lower bound the RHS of~\eqref{eqn: General Inequality 2} as follows
\begin{align}
    & \sum_{t' \in [0 : K]} \sum_{i = \max(0, t' - K + K/G)}^{\min(t', K/G)} \frac{(K/G - i)}{t' + 1} x^{\text{u}}_{t', i} \\
    \geq & \sum_{t' \in [0 : K]} \sum_{i = \max(0, t' - K + K/G)}^{\min(t', K/G)} \frac{(K/G - \min(t', K/G))}{t' + 1} x^{\text{u}}_{t', i} \\
    = & \sum_{t' \in [0 : K]} \frac{(K/G - \min(t', K/G))}{t' + 1} \sum_{i = \max(0, t' - K + K/G)}^{\min(t', K/G)} x^{\text{u}}_{t', i} \\
    = & \sum_{t' \in [0 : K]} G\frac{\binom{K/G}{t' + 1}}{\binom{K/G}{t'}} x^{\text{u}}_{t'}
\end{align}
where $x^{\text{u}}_{t'}$ is defined as
\begin{equation}
    0 \leq x^{\text{u}}_{t'} \coloneqq \sum_{i = \max(0, t' - K + K/G)}^{\min(t', K/G)}  \frac{x^{\text{u}}_{t', i}}{G}.
\end{equation}
After the passages above, we can rewrite~\eqref{eqn: General Inequality 2} as
\begin{equation}
    R_{\text{u}}(\chi) \geq \sum_{t' \in [0 : K]} f_{\text{u}}(t') x^{\text{u}}_{t'}
\end{equation}
where $f_{\text{u}}(t')$ is defined as
\begin{equation}
    f_{\text{u}}(t') \coloneqq G\frac{\binom{K/G}{t' + 1}}{\binom{K/G}{t'}}.
\end{equation}

\subsection{Lower Bounding \texorpdfstring{$R^{\star}$}{Rstar}}
Finally, we can lower bound the optimal worst-case load $R^{\star}$. Indeed, we have the following
\begin{align}
    R^{\star} & \geq \min_{\chi \in \mathcal{X}} \frac{1}{2}\left(R_{\text{c}}(\chi) + R_{\text{u}}(\chi) \right) \\
              & \geq \min_{\chi \in \mathcal{X}} \frac{1}{2} \left(\sum_{t' \in [0 : K]} f_{\text{c}}(t') x^{\text{c}}_{t'} + \sum_{t' \in [0 : K]} f_{\text{u}}(t') x^{\text{u}}_{t'} \right).
\end{align}
Moreover, for any uncoded cache placement $\chi \in \mathcal{X}$ and for some $0 \leq \beta \leq 1$, the following
\begin{align}
    & \sum_{t' \in [0 : K]} x^{\text{c}}_{t'} = 1 \\
    & \sum_{t' \in [0 : K]} t' x^{\text{c}}_{t'} \leq \frac{K\beta M}{\Nc} \label{eqn: Memory-Size Constraint 1}
\end{align}
holds for common files, whereas we have the following
\begin{align}
    & \sum_{t' \in [0 : K]} x^{\text{u}}_{t'} = 1 \\
    & \sum_{t' \in [0 : K]} t' x^{\text{u}}_{t'} \leq \frac{K(1 - \beta) M}{G\Nu} \label{eqn: Memory-Size Constraint 2}
\end{align}
for unique files. This means that we can consider $\bm{x}^{\text{c}} = (x^{\text{c}}_{0}, \dots, x^{\text{c}}_{K})$ and $\bm{x}^{\text{u}} = (x^{\text{u}}_{0}, \dots, x^{\text{u}}_{K})$ as probability mass functions with constraints in~\eqref{eqn: Memory-Size Constraint 1} and~\eqref{eqn: Memory-Size Constraint 2} on the first moment, where such constraints simply represent the maximum memory that is available across the caches of all users for common files and unique files, respectively. In light of the above, we can write
\begin{align}
    R^{\star} & \geq \min_{\chi \in \mathcal{X}} \frac{1}{2} \left(\sum_{t' \in [0 : K]} f_{\text{c}}(t') x^{\text{c}}_{t'} + \sum_{t' \in [0 : K]} f_{\text{u}}(t') x^{\text{u}}_{t'} \right) \\
              & = \min_{\beta, \bm{x}^{\text{c}}, \bm{x}^{\text{u}}} \frac{1}{2} \left(\mathbb{E}_{\bm{x}^{\text{c}}}\left[f_{\text{c}}(t')\right] + \mathbb{E}_{\bm{x}^{\text{u}}}\left[f_{\text{u}}(t') \right] \right) \\
              & \geq \min_{\beta, \bm{x}^{\text{c}}, \bm{x}^{\text{u}}} \frac{1}{2} \left(f_{\text{c}}(\mathbb{E}_{\bm{x}^{\text{c}}}\left[t'\right]) + f_{\text{u}}(\mathbb{E}_{\bm{x}^{\text{u}}}\left[t' \right]) \right) \label{eqn: Application of Jensen's Inequality} \\
              & \geq \min_{\beta} \frac{1}{2} \left(f_{\text{c}}(\tc) + f_{\text{u}}(\tu) \right) \label{eqn: Application of Memory-Size Constraint} \\
              & = \min_{\beta} \frac{1}{2}\left( \frac{\binom{K}{\tc + 1}}{\binom{K}{\tc}} + G \frac{\binom{K/G}{\tu + 1}}{\binom{K/G}{\tu}}\right).
\end{align}
Notice that, since both $f_{\text{c}}(t')$ and $f_{\text{u}}(t')$ are convex and decreasing in $t'$, we have~\eqref{eqn: Application of Jensen's Inequality} and~\eqref{eqn: Application of Memory-Size Constraint} from Jensen's inequality and the constraints on the first moment, respectively. The proof is complete. \qed

\section{Order Optimality Proof}\label{sec: Order Optimality Proof}
From \Cref{thm: Converse Bound Result} we have
\begin{align}
    R^{\star} & \geq \min_{\beta} \frac{1}{2}\left( \frac{\binom{K}{\tc + 1}}{\binom{K}{\tc}} + G \frac{\binom{K/G}{\tu + 1}}{\binom{K/G}{\tu}}\right) \\
              & = \frac{1}{2}\left( \frac{\binom{K}{\tc^{\star} + 1}}{\binom{K}{\tc^{\star}}} + G \frac{\binom{K/G}{\tu^{\star} + 1}}{\binom{K/G}{\tu^{\star}}}\right)
\end{align}
where $\tc^{\star} = K\beta^{\star} M/\Nc$ and $\tu^{\star} = K(1 - \beta^{\star})M/G\Nu$ for some optimal $\beta^{\star}$. Further, from~\cite{Wang2019CodedCachingHeterogeneous} we have
\begin{align}
    R^{\star} & \leq \min_{\beta} \max_{\alpha} R(\beta, \alpha) \\
              & \leq \max_{\alpha} R(\beta^{\star}, \alpha) \label{eqn: Upper Bounding With beta} \\
              & = \max_{\alpha} \frac{\binom{K}{\tc^{\star} + 1} - \binom{G\alpha}{\tc^{\star} + 1}}{\binom{K}{\tc^{\star}}} + G\frac{\binom{K/G}{\tu^{\star} + 1} - \binom{K/G - \alpha}{\tu^{\star} + 1}}{\binom{K/G}{\tu^{\star}}} \\
              & \leq \frac{\binom{K}{\tc^{\star} + 1}}{\binom{K}{\tc^{\star}}} + G\frac{\binom{K/G}{\tu^{\star} + 1}}{\binom{K/G}{\tu^{\star}}}
\end{align}
where the inequality in~\eqref{eqn: Upper Bounding With beta} holds since the optimal value $\beta^{\star}$, which minimizes the lower bound in \Cref{thm: Converse Bound Result}, is not necessarily the optimal memory splitting for the scheme in \Cref{sec: An Existing Achievable Scheme}. To conclude, we have
\begin{equation}
    \frac{1}{2}\left( \frac{\binom{K}{\tc^{\star} + 1}}{\binom{K}{\tc^{\star}}} + G \frac{\binom{K/G}{\tu^{\star} + 1}}{\binom{K/G}{\tu^{\star}}}\right) \leq R^{\star} \leq \frac{\binom{K}{\tc^{\star} + 1}}{\binom{K}{\tc^{\star}}} + G \frac{\binom{K/G}{\tu^{\star} + 1}}{\binom{K/G}{\tu^{\star}}}
\end{equation}
which implies that the coded scheme in \Cref{sec: An Existing Achievable Scheme} is order optimal within a constant multiplicative factor of $2$. The proof is complete. \qed

\section{Conclusion}\label{sec: Conclusion}

In this paper, we considered a coded caching setting with heterogeneous user profiles. Under the system model originally proposed in~\cite{Wang2019CodedCachingHeterogeneous}, we constructed a novel information-theoretic converse on the worst-case communication load under uncoded prefetching. We developed the lower bound by taking advantage of the genie-aided approach introduced in~\cite{Yu2018ExactRateMemory}. Interestingly, the proposed converse bound, jointly with the Scheme 2 from~\cite{Wang2019CodedCachingHeterogeneous}, allows us to characterize the optimal worst-case load under uncoded prefetching within a constant multiplicative factor of $2$. Although the converse in \Cref{thm: Converse Bound Result} holds under the constraint of uncoded placement, the result in \Cref{thm: Order Optimality Result}, which provides a constant order optimality factor independent of all system parameters, improves the previously known order optimality results in~\cite{Zhang2022CodedCachingHeterogeneous}. Possible extensions could include the study of other (maybe more complex) heterogeneous user profiles as well as establishing the exact fundamental limits of the setting considered in this paper.

\bibliographystyle{IEEEtran}
\bibliography{references}

\end{document}